\documentclass[jgrga]{agutex} 

  \usepackage[dvips]{graphicx}
 \setkeys{Gin}{draft=false}
\usepackage{amsmath,amssymb}

\authorrunninghead{MURPHY ET AL.}

\titlerunninghead{ASYMMETRIC MAGNETIC RECONNECTION}

\authoraddr{P.~A.~Cassak,
Department of Physics, Hodges Hall, Box 6315,
West Virginia University,
Morgantown, WV 26506, USA.} 

\authoraddr{N.~A.~Murphy,
Harvard-Smithsonian Center for Astrophysics,
60 Garden Street, MS 16, 
Cambridge, MA 02138, USA.
(namurphy@cfa.harvard.edu)}

\authoraddr{C.~R.~Sovinec,
Department of Engineering Physics,
University of Wisconsin, 
1500 Engineering Drive,
Madison, WI 53706, USA.}

\newcommand{\xhat}{\ensuremath{\hat{\mathbf{x}}}}
\newcommand{\yhat}{\ensuremath{\hat{\mathbf{y}}}}
\newcommand{\zhat}{\ensuremath{\hat{\mathbf{z}}}}
\newcommand{\ihat}{\ensuremath{\boldsymbol{\hat{\mathsf{I}}}}} 

\newcommand{\dif}{\ensuremath{\mathrm{d}}}

\newcommand{\e}{\ensuremath{\mathrm{e}}}

\begin{document}

%
%

\newcommand{\sech}{\ensuremath{\rm{sech}}}

\newcommand{\ltot}{\ensuremath{L}}

\title{Magnetic Reconnection with Asymmetry in the Outflow Direction} 
\date{\today} 

%
%






\authors{N. A. Murphy, \altaffilmark{1,2,3}
C. R. Sovinec, \altaffilmark{1,4} and
P. A. Cassak \altaffilmark{5}}

\altaffiltext{1}{Center for Magnetic Self-Organization in Laboratory
and Astrophysical Plasmas, University of Wisconsin, Madison,
Wisconsin, USA.}
\altaffiltext{2}{Department of Astronomy, University of Wisconsin,
  Madison, Wisconsin, USA.}
\altaffiltext{3}{Harvard-Smithsonian Center for Astrophysics,
  Cambridge, Massachusetts, USA.}
\altaffiltext{4}{Department of Engineering Physics, University of
  Wisconsin, Madison, Wisconsin, USA.}
\altaffiltext{5}{Department of Physics, West Virginia University,
  Morgantown, West Virginia, USA.}

%
%


\begin{abstract}
Magnetic reconnection with asymmetry in the outflow direction occurs
in the Earth's magnetotail, coronal mass ejections, flux cancellation
events, astrophysical disks, spheromak merging experiments, and
elsewhere in nature and the laboratory.  A control volume analysis is
performed for the case of steady antiparallel magnetic reconnection
with asymmetric downstream pressure, which is used to derive scaling
relations for the outflow velocity from each side of the current sheet
and the reconnection rate.  Simple relationships for outflow velocity
are presented for the incompressible case and the case of symmetric
downstream pressure but asymmetric downstream density.  Asymmetry
alone is not found to greatly affect the reconnection rate.  The flow
stagnation point and magnetic field null do not coincide in a steady
state unless the pressure gradient is negligible at the flow
stagnation point.
\end{abstract}

\begin{article}

\section{Introduction}

While most two-dimensional models of magnetic reconnection assume that
the process is symmetric to a 180$^\circ$ rotation about the X-point,
there are many situations in nature and in the laboratory where this
assumption is invalid.  In recent years, many papers have addressed
magnetic reconnection with asymmetry in the inflow direction
\citep[e.g.,][]{ lol:asymmetric, nakamura:2000, swisdak:diamagnetic,
oieroset:2004, borovsky:asymmetric, cassak:asymmetric, cassak:hall,
cassak:dissipation, birn:asymmetric, murphy:mrx, mozer:asym,
pritchett:asym, borovsky:2008, tanaka:2008, mozer:pritchett:2009,
eriksson:2009}.  In particular, \citet{cassak:asymmetric} generalized
the Sweet-Parker model \citep{parker,sweet} to account for
reconnection between plasmas with different upstream densities and
magnetic field strengths.  They found that the reconnection rate is
governed by a hybrid Alfv\'en speed that takes into account the
densities and magnetic field strengths for the two upstream regions.
The positions of the magnetic field null and flow stagnation point are
displaced from each other, with the field null position set by balance
of energy flux and the stagnation point position set by balance of
mass flux.  In addition to reconnection with asymmetric inflow, there
are many situations in nature and the laboratory for which the system
is asymmetric in the outflow direction.  In this paper, we analyze
steady magnetic reconnection with asymmetry in the outflow direction.

The best known scenario for magnetic reconnection with asymmetry in
the outflow direction is the Earth's magnetotail.  In this case,
asymmetry is a particularly important consideration because it helps
determine the amount of energy transported in the earthward and
tailward directions as a result of reconnection.  At distances of
$\sim$5--15$R_E$, there is a considerable pressure gradient as the
plasma pressure decreases approximately monotonically with distance
from Earth \citep{lui:quiet,shiokawa:braking,xing:2009}.
Earthward-directed reconnection outflow must work against strong
gradients in both plasma pressure and magnetic pressure.  Because of
the global configuration of the magnetotail, the X-line
characteristically moves in the tailward direction \citep{hones:1979}.
Reconnection with asymmetry in the outflow direction has often been
seen in simulations of the magnetotail \citep[e.g.,][]{birn:1996,
hesse:1996, hesse:2001, kuznetsova:2007, laitinen:2005,
laitinen:thesis, birn:2009, zhu:2009}, though the degree of asymmetry
depends on the proximity of the reconnection layer to Earth and how
reconnection is driven.  The largest discrepancy between earthward and
tailward outflow velocities in these simulations was seen by
\cite{laitinen:2005} and \cite{laitinen:thesis}, where the inflow had
a large component of velocity in the outflow direction; consequently,
there is a large separation between the X-line and the flow reversal
line in their results.

Observing reconnection with asymmetry in the outflow direction in the
magnetosphere requires multiple satellites crossing the earthward and
tailward sides of the diffusion region at approximately the same
time. While statistical approaches are possible
\citep[e.g.,][]{2009JGRA..11409203P}, observations of a single event
are not common.  One occurrence is a crossing of the diffusion region
by Cluster on 11 October 2001.  Cluster was in the region between the
outflow jets from 03:30--03:36 UT, but passed the X-line at 03:31
UT\@.  One possible explanation of this is that the X-line was near
the tailward end of the diffusion region.  However, other explanations
(e.g., time-dependent behavior or undetected additional X-lines)
cannot be ruled out with the available data \citep{laitinen:2007}.

In solar physics, reconnection during coronal mass ejections (CMEs),
solar flares, and flux cancellation events are asymmetric in the
outflow direction when one outflow jet is directed sunward and the
other outflow jet is directed away from the Sun
\citep[e.g.,][]{kopp:1976, martin:1985, shibata:1995, litvinenko:1999,
linforbes:2000, 2002A&A...384..273A}.  Observations of bidirectional
jets in the solar atmosphere \citep[see][]{innes:1997, wang:2007,
liu:2009, 2009A&A...506L..45G} show that, despite the effects of
gravity, the redshifted jet is often slower than the blueshifted jet
because the redshifted jet must propagate into a higher density
medium.  In these events, gravity's most important effect is the
establishment of a stratified medium.  Current sheets forming in such
a medium are likely to have strong gradients in the outflow direction
for upstream density, pressure, and magnetic field strength
\citep[see][]{2002ApJ...575.1116C, 2003ApJ...594.1068K,
2004ApJ...602L..61C, 2006ApJ...638.1110B, 2007ApJ...658L.123L,
2008ApJ...686.1372C, 2008ApJ...689..572B, 2009ApJ...693.1666L,
2009A&A...499..905V, 2009ApJ...699..245S, 2009A&A...506..901A}.
Simulations of reconnection in a stratified medium show that the
redshifted jet can be up to an order of magnitude slower than the
blueshifted jet \citep{roussev:2001}, and that reconnection in such an
atmosphere displays a more complicated velocity structure than
symmetric two-dimensional reconnection \citep{galsgaard:2002}.
Gravity itself can be an important consideration if the work done by
electromagnetic forces is comparable to or less than the work done
against gravity \citep{reeves06}.  Asymmetry in the outflow direction
also happens when magnetic field lines in one downstream region are
line-tied while magnetic field lines in the other downstream region
are open.

During turbulent reconnection \citep[e.g.,][]{lazarian:turbrecon} and
reconnection occurring during a turbulent cascade
\citep[e.g.,][]{servidio:turbrecon}, there will in general be many
reconnection sites throughout the volume of interest.  Reconnection
occurring at each of these sites will in general be asymmetric in the
inflow and outflow directions, as well as the out-of-plane direction.
Reconnection processes involving multiple competing reconnection sites
or multiple magnetic islands \citep[e.g.,][]{1986JGR....91.6807L,
drake:nature, lin:2008,chen:2009} will also likely involve asymmetry
in the outflow direction, especially if the X-lines are not evenly
spaced or develop at different rates.

In astrophysical settings, the winds of strongly magnetized hot stars
(e.g., the Bp star $\sigma$~Ori~E) can be channeled along a
predominantly dipolar field to form an equatorial circumstellar disk
or buildup of material \citep{nakajima:sigmaoriE, cassinelli:mtd,
townsend:rrm}.  While the dipole field is in general dominant close to
the star, recent axisymmetric simulations show that the continual
funneling of material can eventually lead to centrifugal breakout
events associated with magnetic reconnection
\citep{uddoula:centrifugal,uddoula:2008}.  In this case, the
reconnection outflow is aligned with the radial direction, with one
exhaust path directed towards the disk and the star, and the other
leading to the interstellar medium.  Such reconnection events could be
the source of the X-ray flares observed on $\sigma$~Ori~E by ROSAT
\citep{groote:2004}.  Considerations of asymmetry in the outflow
direction are also important for magnetic reconnection events
associated with centrifugal instabilities and plasma release in the
Jovian magnetosphere \citep[e.g.,][]{kivelson:2005}.

In the laboratory, reconnection with asymmetry in the outflow
direction occurs during the merging of spheromaks and in toroidal
plasma configurations where the reconnection outflow is aligned with
the radial direction.  Relevant experiments include the Swarthmore
Spheromak Experiment (SSX) \citep{cothran:ssx}, the Magnetic
Reconnection Experiment (MRX) \citep{yamada:mrx}, and TS-3/4 at the
University of Tokyo \citep{ono:1993}.  Recent spheromak merging
experiments at MRX have shown that asymmetry in the outflow direction
develops as a result of the Hall effect \citep{inomoto:counter}.  In
these experiments at MRX, the reconnecting magnetic field lines do not
lie in the poloidal plane, and there is a component of the electron
flow associated with the reconnection current in the radial
direction. This radial component of electron velocity pulls the
reconnecting field lines, leading to a shift in position of the
X-point, asymmetric outflow, and asymmetric downstream pressure.
Reversing the toroidal field direction changes the direction of the
shift, but because of toroidicity, this also changes the reconnection
rate and radial pressure profile \citep{inomoto:counter,murphy:mrx}.
Recent simulations of spheromak merging in SSX show reconnection with
much stronger radially inward directed outflow even though the plasma
pressure near $R=0$ is large due to a pileup of exhaust
\citep{lin:ssx}.  These results suggest that considerations of
asymmetry in the outflow direction are important for the
interpretation of bidirectional jets recently reported in experiment
\citep{brown:2006}.

\citet{murphy:mrx} presented simulations of the reconnection process
in the geometry of MRX, showing that asymmetric inflow occurs during
the pull mode of operation and asymmetric outflow during the push mode
of operation \citep[see][Figure 3]{yamada:mrx}.  The inboard (low
radius) side of the current sheet is more susceptible to buildup or
depletion of density due to the lesser available volume than on the
outboard (high radius) side of the current sheet.  As a result of the
pressure buildup at low radii during push reconnection, the X-point is
located closer to the outboard side of the current sheet than the
inboard side.  Consequently, the radially inward directed outflow is
subjected to a stronger tension force than the radially outward
directed outflow, allowing comparable outflow velocities from both the
inboard and outboard sides of the current sheet (a similar effect is
discussed by \citet{galsgaard:2002}).  During several time intervals
in these simulations and despite the higher pressure in the inboard
downstream region, the radially inward directed outflow speed is found
to be greater than the radially outward directed outflow speed.  The
magnetic field null and flow stagnation point are separated during
both pull and push reconnection \cite[][Figures 2.4 and
2.6]{murphy:thesis}.  Push reconnection is an example of how asymmetry
in the outflow direction develops when outflow in one downstream
region is confined more effectively than outflow in the other
downstream region.

\citet{oka:asym} performed particle-in-cell (PIC) simulations of
reconnection where outflow from one end of the current sheet is
impeded by a hard wall while outflow from the other end encounters no
such obstruction.  They found that the X-line retreats from the wall
at {$\sim$}{$10$}\% of the upstream Alfv\'en velocity $V_A$ and that
the reconnection rate is largely unchanged from the symmetric case.
Moreover, there is a separation between the ion flow stagnation point
and the magnetic field null, with the field null located further from
the wall than the ion flow stagnation point.  In a work that relates
asymmetry in the inflow direction with asymmetry in the outflow
direction, \citet{swisdak:diamagnetic} found that the presence of a
density gradient in the inflow direction across a current sheet can
lead to a drift of the X-line in the electron diamagnetic drift
direction when a guide field is present \citep[see
also][]{rogers:1995}.  The reconnection process is suppressed when the
drift velocity is comparable to or greater than the Alfv\'en velocity.
The effects of current sheet motion and time-dependence on slow shock
mediated reconnection layers have also been considered
\citep{owen:1987:451,owen:1987:467,kiehas:2007,kiehas:2009}.

In this paper, we perform a control volume analysis for a current
sheet with asymmetric downstream pressure and test the resulting
scaling relations against simulations.  The objectives are to
determine (1) the relationship between the upstream parameters, the
downstream pressures, and the reconnection outflow velocity, (2) how
the reconnection rate is affected by asymmetric downstream pressure,
and (3) what sets the positions of the magnetic field null and flow
stagnation point.  In section \ref{equations}, we write the equations
of resistive magnetohydrodynamics (MHD) in a time-independent integral
form that is amenable to a control volume analysis.  In section
\ref{lineargeom}, we review the effects of symmetric downstream
pressure on antiparallel reconnection and develop scaling relations
for a current sheet with asymmetric downstream pressure.  In section
\ref{sim}, we test the scaling relations derived in section
\ref{lineargeom} against resistive MHD simulations of reconnection
with asymmetry in the outflow direction.  In section
\ref{conclusions}, we provide a discussion and summarize our results.
A similar analysis for reconnection in cylindrical geometry with
outflow aligned with the radial direction is presented by
\citet[][section 3.4]{murphy:thesis}.

\section{Equations of Magnetohydrodynamics\label{equations}}

The equations of resistive MHD in conservative form \citep[e.g.,][pp.\
165--166]{goedbloed:poedts} are
\begin{eqnarray}
  \frac{\partial\rho}{\partial t} + 
  \nabla\cdot\left(\rho\mathbf{V}
  \right)=0,  \label{mass_cons}
\\
  \frac{\partial (\rho\mathbf{V})}{\partial t} + 
  \nabla\cdot\left[
    \rho \mathbf{VV} + 
    \left( p + \frac{B^2}{2\mu_0} \right)\ihat -
    \frac{\mathbf{BB}}{\mu_0}
  \right]=0, \label{momentum_cons}
\\
  \frac{\partial\mathcal{E}}{\partial t} +
  \nabla\cdot\left[
    \left(
      \frac{\rho V^2}{2} + \frac{\gamma}{\gamma-1}p
    \right)\mathbf{V}
    + \frac{\mathbf{E}\times\mathbf{B}}{\mu_0}
  \right]=0, \label{energy_cons} 
\\
  \frac{\partial\mathbf{B}}{\partial t} + 
  \nabla\times\mathbf{E}\label{faraday_cons}=0,
\\
  \mu_0\mathbf{J} = \nabla\times\mathbf{B},\label{ampere_cons}
\\
  \mathbf{E} + \mathbf{V}\times\mathbf{B} = \eta\mathbf{J},
\label{ohms_cons}
\end{eqnarray}
where $\mathbf{B}$ is the magnetic field, $\mathbf{E}$ is the electric
field, $\mathbf{V}$ is the bulk plasma velocity, $\mathbf{J}$ is the
current density, $p$ is the plasma pressure, $\rho$ is mass density,
$\eta$ is the plasma resistivity, $\mathcal{E}\equiv \rho V^2/2 +
p/(\gamma-1) + B^2/2\mu_0$ is the total energy density, and $\gamma$
is the ratio of specific heats.  The identity dyadic tensor is given
by $\ihat=\xhat\xhat+\yhat\yhat+\zhat\zhat$.  Equation
(\ref{energy_cons}) includes the the internal energy flux,
$p\mathbf{V}/(\gamma-1)$, and the mechanical work done on or by the
plasma by pressure gradients while moving, $p\mathbf{V}$.

Following the approach presented by \citet{cassak:asymmetric}, we
assume a steady-state system, integrate over an arbitrary closed
volume $\mathcal{V}$ bounded by the surface $\mathcal{S}$, and use the
divergence theorem to write the continuity, momentum, and energy
equations as
\begin{eqnarray}
\oint_{\mathcal{S}} \dif\mathbf{S}\cdot 
 \left( \rho\mathbf{V}  \right)=0,  \label{mass_int}
\\
\oint_{\mathcal{S}} \dif\mathbf{S}\cdot
\left[
\rho\mathbf{VV} + 
\left(
p + \frac{B^2}{2\mu_0}
\right)\ihat
- \frac{\mathbf{BB}}{\mu_0}
\right]=0, \label{momentum_int}
\\
\oint_{\mathcal{S}} \dif\mathbf{S}\cdot
\left[
\left( \frac{\rho V^2}{2} + \frac{\gamma p}{\gamma-1}\right)\mathbf{V}
+ \frac{\mathbf{E}\times\mathbf{B}}{\mu_0} 
\right]=0,\label{energy_int}
\end{eqnarray}
where $\dif\mathbf{S}$ is a differential area element pointing in the
outward normal direction to $\mathcal{S}$.  Similarly, with the help
of Stokes' theorem, equation (\ref{faraday_cons}) leads to
\begin{equation}
\oint_{\mathcal{S}} \dif\mathbf{S}\times\mathbf{E}=0.\label{faraday_int}
\end{equation}
Equations (\ref{mass_int})--(\ref{faraday_int}) are valid for any
arbitrary closed volume, provided a steady-state has been achieved.
These surface integrals are evaluated in section \ref{lineargeom} to
investigate magnetic reconnection with asymmetry in the outflow
direction.

\section{Scaling Relations\label{lineargeom}}

The Sweet-Parker model \citep{sweet,parker} describes symmetric
steady-state antiparallel magnetic reconnection in the resistive MHD
framework when compressibility, viscosity, and downstream pressure are
unimportant.  In this section, we extend these results to account for
reconnection with asymmetric downstream pressure.  After reviewing the
effects of symmetric downstream pressure on the reconnection process
in subsection \ref{symmetric}, we consider the case of asymmetric
downstream plasma pressure in subsection \ref{asymmetric}.  We then
investigate the internal structure of such a current sheet in
subsection \ref{internal}.

\subsection{Effects of Symmetric Downstream Pressure\label{symmetric}}

The effects of symmetric downstream pressure on a Sweet-Parker current
sheet are discussed by \citet[][pp.\ 123--126]{PF00}.  Presently, we
review their results using the approach that we employ later this
section for a current sheet with asymmetric downstream pressure while
relaxing their assumptions regarding compressibility \citep[see
also][]{parker:1963,chae:2003,litvinenko:2009}.  The characteristic
parameters used in this derivation are: $B_{in}$, upstream magnetic
field strength; $V_{in}$, plasma inflow velocity; $V_{out}$, plasma
outflow velocity; $p_{in}$, upstream plasma pressure; $p_{out}$,
downstream plasma pressure; $\rho_{in}$, upstream plasma density;
$\rho_{out}$, downstream plasma density; $J_y$, out-of-plane current
density inside the layer; $E_y$, out-of-plane electric field; $\ltot$,
current sheet half-length; and $\delta$, current sheet half-thickness.
We define $x$ as the outflow direction, $y$ as the out-of-plane
direction, and $z$ as the inflow direction.

Everywhere except within the reconnection layer, the ideal Ohm's law
is approximately valid.  By assuming a steady state the electric field
is constant and given by
\begin{equation}
E_y = V_{in}B_{in} \label{fluxcon}.
\end{equation}
Since $B_x$ reverses over a distance of $\sim$2$\delta$, Ampere's law
gives
\begin{equation}
J_y \sim \frac{B_{in}}{\mu_0\delta}\label{ampere_scale}.
\end{equation}
Matching the resistive electric field inside the layer with the ideal
electric field outside the layer gives
\begin{equation}
V_{in} \sim \frac{\eta}{\mu_0\delta}\label{inflow_whole}.
\end{equation}

Evaluating the conservation of mass relation given in equation
(\ref{mass_int}) over the entire volume of the current sheet yields
the relation
\begin{equation}
  \rho_{in}V_{in}\ltot \sim \rho_{out}V_{out}\delta. \label{mass_sym}
\end{equation}
The conservation of momentum surface integral given in equation
(\ref{momentum_int}) is satisfied by any distribution of fluxes with
the assumed symmetry when integrating over the outer boundary of the
current sheet.  Evaluating the conservation of energy relation given
in equation (\ref{energy_int}) yields the relation
\begin{equation}
  V_{in}\ltot\left( \alpha p_{in} + \frac{B_{in}^2}{\mu_0} \right) \sim
  V_{out}\delta \left( \alpha p_{out}+\frac{\rho_{out} V_{out}^2}{2}\right),
  \label{energy_sym}
\end{equation}
where $\alpha \equiv \gamma/(\gamma-1)$.  Here we neglect
contributions from upstream kinetic energy and downstream magnetic
energy.  Dividing equation (\ref{energy_sym}) by equation
(\ref{mass_sym}) and rearranging gives the scaling relation
\begin{equation}
  V_{out}^2 \sim 
   V_A^2 -
   \alpha \left(
    \frac{p_{out}}{\rho_{out}} - \frac{p_{in}}{\rho_{in}}
    \right),
  \label{outflow_sym}
\end{equation}
where $V_A\equiv B_{in}/\sqrt{\mu_0\rho_{in}}$ is the upstream
Alfv\'en speed and we ignore factors of order unity.  The term $\alpha
p/\rho$ is the enthalpy per unit mass.  Using equation
(\ref{inflow_whole}), the scaling for the dimensionless reconnection
rate can then be written as
\begin{equation}
  \frac{V_{in}}{V_A} \sim
  \frac{1}{S^{1/2}}
  \sqrt{\frac{\rho_{out}}{\rho_{in}}}
  \left[
    1 - \frac{\alpha}{V_A^2}
    \left(
      \frac{p_{out}}{\rho_{out}}-\frac{p_{in}}{\rho_{in}}
    \right)
  \right]^{1/4}
  \label{symmetric_recon_rate}
\end{equation}
where $S\equiv \mu_0 \ltot V_A/\eta$ is the Lundquist number.  The
reconnection rate depends weakly on the downstream pressure except
when the bracketed quantity is close to zero.  The Sweet-Parker
scalings of $V_{out}\sim V_A$ and $V_{in}/V_A \sim S^{-1/2}$ are
recovered when $\rho_{out}/\rho_{in}$ and the quantity in brackets are
independent of $S$.  We also see that when compression makes the
outflow density larger than the inflow density, it relaxes the usual
bottleneck from flow moving through the reconnection region.

\subsection{Effects of Asymmetric Downstream Pressure\label{asymmetric}}

\begin{figure}
\includegraphics[scale=1]{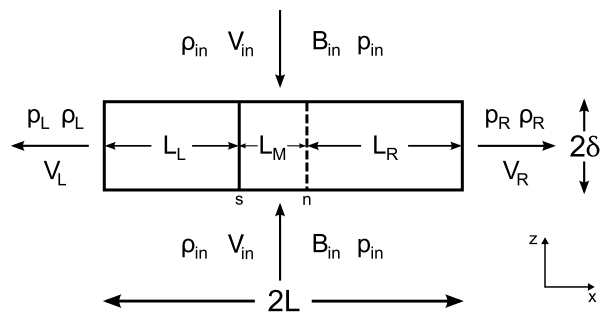}
\caption{Sweet-Parker-like reconnection with asymmetric downstream
  pressure and a pressure gradient through the current sheet.  The
  solid vertical bar inside the current sheet (marked {s}) represents
  the flow stagnation point and the dashed vertical bar (marked {n})
  represents the magnetic field null.}
  \label{asym_diagram} 
\end{figure}

We now consider a current sheet with symmetric inflow but with
asymmetric outflow and downstream pressure.  In this framework, it is
necessary to assume that the current sheet position and structure is
steady within the inertial reference frame of the X-line.  For
example, reconnection could be externally driven in such a way that
constrains the position of the current sheet.  The setup of this
problem is shown in Figure \ref{asym_diagram}.  Throughout this
analysis, subscripts $L$ and $R$ indicate that the variable represents
the characteristic downstream value of a field for the left and right
sides of the current sheet.

To proceed, we evaluate the surface integrals given in equations
(\ref{mass_int})--(\ref{energy_int}) over the whole volume of the
current sheet depicted in Figure \ref{asym_diagram}.  The conservation
of mass surface integral given in equation (\ref{mass_int}) yields the
relation
\begin{equation}
  2\ltot \rho_{in}V_{in} \sim \rho_LV_L\delta +
  \rho_RV_R\delta.\label{mass_whole} 
\end{equation}

Evaluating the component of the conservation of momentum surface
integral given in equation (\ref{momentum_int}) in the outflow
direction yields a relation between the plasma pressures and momentum
fluxes from each exit of the reconnection layer,
\begin{equation}
  \rho_L V_L^2 + p_L \sim \rho_R V_R^2 + p_R.\label{momentum_whole}
\end{equation}
Because the current sheet is assumed to be long and thin, the above
relation neglects forces due to the downstream magnetic field.
However, magnetic tension does not need to be negligible throughout
the volume of integration for this relationship to hold.  Rather,
tension need only either be negligible along the boundary or
contribute along the boundary evenly in both outflow directions.  If
the upstream magnetic field is not parallel to the boundaries along
$z=\pm\delta$ in a way which is not symmetric in the outflow
direction, this may yield an additional contribution by tension
towards momentum balance in the outflow direction.  Downstream
magnetic pressure can be important when the global magnetic field
configuration contains a large vertical component that impedes outflow
from one side of the current sheet \citep[e.g.,][Figure
5]{inomoto:counter}.  We also assume that the momentum flux
$\rho\mathbf{VV}$ into the current sheet does not significantly
contribute to momentum balance in the outflow direction; this is
expected to be important only when the outflow component of the inflow
velocity is of the same order as the outflow velocities.  Force
balance must be met in both the inflow and outflow directions
simultaneously in order for the assumption of time-independence to be
valid.

Using the expression for the electric field given in equation
(\ref{fluxcon}), the energy conservation integral (\ref{energy_int})
provides the relation
\begin{eqnarray}
  2\ltot V_{in}\left( \alpha p_{in} + \frac{B_{in}^2}{\mu_0}\right) \sim
  V_L\delta\left(\alpha p_L +\frac{\rho_L V_L^2}{2} \right) +
  \nonumber\\
  V_R\delta\left( \alpha p_R + \frac{\rho_R V_R^2}{2} 
  \right).\label{energy_whole} 
\end{eqnarray}
The above relation neglects upstream kinetic energy and the Poynting
flux out of the layer.

By using equation (\ref{mass_whole}) to eliminate $2\ltot V_{in}$ from
equation (\ref{energy_whole}) and equation (\ref{momentum_whole}) to
eliminate $V_R$, we arrive at the following cubic relationship which
can be solved for $V_L^2$,
\begin{equation}
  0 \sim C_{6L}V_L^6 + C_{4L}V_L^4 +  C_{2L}V_L^2 + C_{0L},
  \label{linear_cubic}
\end{equation}
where we do not explicitly assume the nature of the dissipation
mechanism.  The coefficients for the above equation are functions of
the upstream magnetic field strength as well as the upstream and
downstream densities and pressures, and are given by
\begin{eqnarray}
  C_{6L} &\equiv&
  \frac{1}{4}
  \left(
    \frac{\rho_L^3}{\rho_R} - \rho_L^2
  \right),\label{C6L}\\
  C_{4L} &\equiv&
  \frac{\rho_L^2}{\rho_R}
  \left(
    \alpha p_R - \frac{3}{4}\Delta p
  \right)
  - \alpha \rho_L p_L  \\
  C_{2L}&\equiv &
  \rho_L\left(\rho_R-\rho_L\right)c_{in}^4
  + 2\rho_L\Delta p\left(1-\alpha\right)c_{in}^2- \alpha^2 p_L^2\nonumber\\ & &
    + \alpha^2 p_R^2 
    \left[
      \frac{\rho_L}{\rho_R}   
      \left( 1 - \frac{ \Delta p}{2\alpha p_R} \right)
      \left( 1 - \frac{3\Delta p}{2\alpha p_R} \right)
    \right]
  \\
  C_{0L} &\equiv& - \rho_R \Delta p
  \left[
    c_{in}^2 - \frac{1}{2}
    \left(
      \frac{2\alpha p_R-\Delta p}{\rho_R}
    \right)
  \right]^2
  ,\label{C0L}
\end{eqnarray}
where the velocity $c_{in}$ is defined as
\begin{equation}
  c_{in}^2 \equiv \frac{B_{in}^2}{\mu_0\rho_{in}} 
  + \alpha\frac{p_{in}}{\rho_{in}},\label{cin}
\end{equation}
and we define the average downstream pressure $\bar{p}$ and the
pressure difference $\Delta p$ as
\begin{eqnarray}
  \bar{p} & \equiv & \frac{p_L+p_R}{2},  \label{pbar} \\
  \Delta p & \equiv &  p_R - p_L. \label{deltap}
\end{eqnarray}
Equation (\ref{linear_cubic}) was derived assuming that the scaling
factors given in equations (\ref{mass_whole}), (\ref{momentum_whole}),
and (\ref{energy_whole}) are unity.  If this is not the case, then if
$\xi$ is equal to the right hand side divided by the left hand side of
equation (\ref{mass_whole}), and $\zeta$ is equal to the right hand
side divided by the left hand side of equation (\ref{energy_whole}),
then the transformation $c_{in}^2\rightarrow \left(\zeta/\xi\right)
c_{in}^2$ will algebraically account for scaling factors that are not
unity in equations (\ref{mass_whole}) and (\ref{energy_whole}) for
equation (\ref{linear_cubic}).

Equation (\ref{linear_cubic}) simplifies for some special cases.  When
$\rho_L=\rho_R\equiv\rho_{out}$, the coefficient $C_{6L}$ vanishes,
leaving a quadratic equation in $V_L^2$.  In the incompressible limit
with $\rho_{in}=\rho_L=\rho_R\equiv\rho$ and $\alpha=1$, the solution
becomes
\begin{equation}
  V_{L,R}^2 \sim
  \sqrt{
    4\left(
      c_{in}^2 - \frac{\bar{p}}{\rho}
    \right)^2
    +
    \left(
      \frac{\Delta p}{2\rho}
    \right)^2
  }
  \pm \frac{\Delta p}{2\rho},\label{incomp}
\end{equation}
where the plus and minus signs refer to $V_L$ and $V_R$, respectively.
This gives the expected result that the outflow speed is slower on the
side with higher downstream pressure.

Next, consider the special case with $p_L=p_R=p_{out}$, but where the
downstream densities can be different.  In this case, $C_{0L}$
vanishes, again leaving a quadratic equation.  The solution is
\begin{equation}
  V_{L,R}^2 \sim  c_{in}^2 \sqrt{\frac{\rho_{R,L}}{\rho_{L,R}}}
  - \alpha \frac{p_{out}}{\rho_{L,R}},\label{isobaric}
\end{equation}
where we ignore factors of order unity.  This equation shows that the
outflow speed is higher on the low density side.  The scalings
presented in equations (\ref{incomp}) and (\ref{isobaric}) both reduce
to the scaling in equation (\ref{outflow_sym}) in the symmetric limit
as they should.

Equations (\ref{linear_cubic}), (\ref{incomp}), and (\ref{isobaric})
are derived solely from the scaling relations for conservation of
mass, energy, and momentum without explicitly assuming a dissipation
mechanism \citep[see also][]{cassak:asymmetric}.  By using equations
(\ref{inflow_whole}) and (\ref{mass_whole}) (and consequently assuming
uniform resistive dissipation inside the current sheet), the inflow
speed can be written as
\begin{equation}
  \frac{V_{in}}{V_A} \sim 
  \sqrt{
    \frac{V_L+V_R}{2V_AS}
  }.\label{asym_recon_rate}
\end{equation} 
Using equations (\ref{fluxcon}) and (\ref{asym_recon_rate}), the
reconnection rate is
\begin{equation}
  E_y \sim B_{in}
  \sqrt{\frac{\eta\left(V_L+V_R\right)}{2\mu_0L}}
  \label{reconEy}.
\end{equation}

\begin{figure}
  \includegraphics{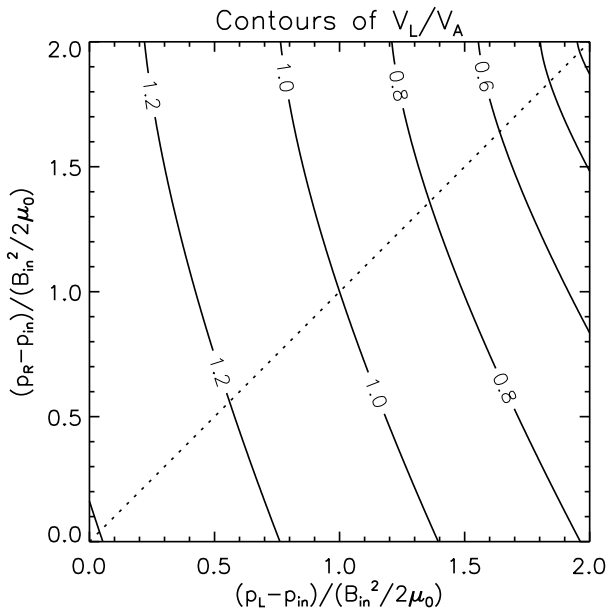}
  \caption{ Contours of the outflow velocity from the left side of the
    current sheet, $V_L/V_A$, calculated from equation (\ref{incomp})
    as a function of $p_L-p_{in}$ and $p_R-p_{in}$ for the
    incompressible case.  Contours are separated by $0.2$.  Figures
    \ref{VL_contour}, \ref{norm_recon_rate}, and \ref{plot1d} assume
    that the scaling factors for equations (\ref{mass_whole}),
    (\ref{momentum_whole}), and (\ref{energy_whole}) are unity.  }
  \label{VL_contour}
\end{figure}


\begin{figure}
  \includegraphics{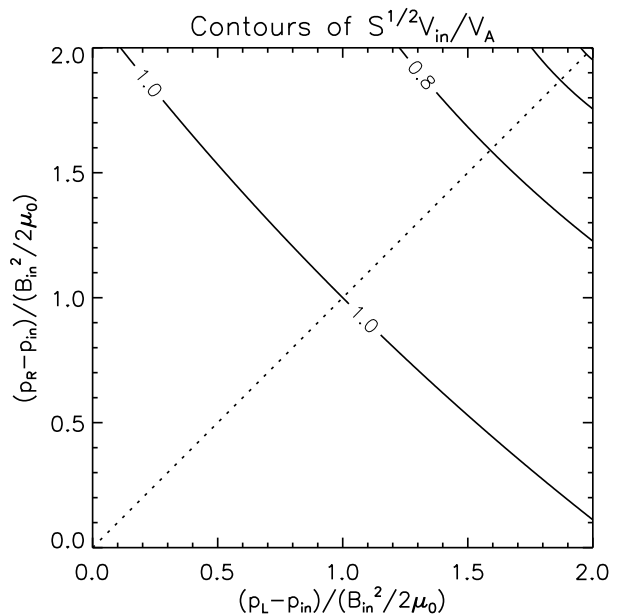}
  \caption{ Contours of the normalized reconnection rate, given by
    $S^{1/2}V_{in}/V_A = \sqrt{(V_L+V_R)/2V_A}$, as a function of
    $p_L-p_{in}$ and $p_R-p_{in}$ and calculated using equation
    (\ref{incomp}) to find $V_L$ and $V_R$ for the incompressible
    case.  Contours are separated by $0.2$.  }
  \label{norm_recon_rate}
\end{figure}

Solutions of equation (\ref{incomp}) for $V_L$ as a function of
$p_L-p_{in}$ and $p_R-p_{in}$ in units of the upstream magnetic
pressure are presented in Figure \ref{VL_contour} for the
incompressible limit.  The value for $V_R$ can be found by switching
the values for $p_L-p_{in}$ and $p_R-p_{in}$.  We see that the outflow
velocity from one end does not depend strongly on the downstream
pressure from the opposite end.  In fact, reconnection events (e.g.,
in the solar atmosphere) do not require bidirectional outflow jets
traveling at the Alfv\'en speed.  Rather, in the presence of
asymmetric downstream pressure, there can be one Alfv\'enic jet and
one sub-Alfv\'enic jet \citep[see also][]{roussev:2001}.  As shown by
the widely spaced contours in Figure \ref{norm_recon_rate}, the
normalized reconnection rate $S^{1/2}V_{in}/V_A=\sqrt{(V_L+V_R)/2V_A}$
is only weakly dependent on the difference in downstream pressures.
This conclusion is consistent with the simulations of X-line retreat
reported by \citet{oka:asym}, in which the reconnection rate is not
greatly affected when outflow from one reconnection jet is impeded by
the presence of an obstacle while the other outflow jet has no such
obstruction.

\begin{figure}
  \begin{center}
    \includegraphics{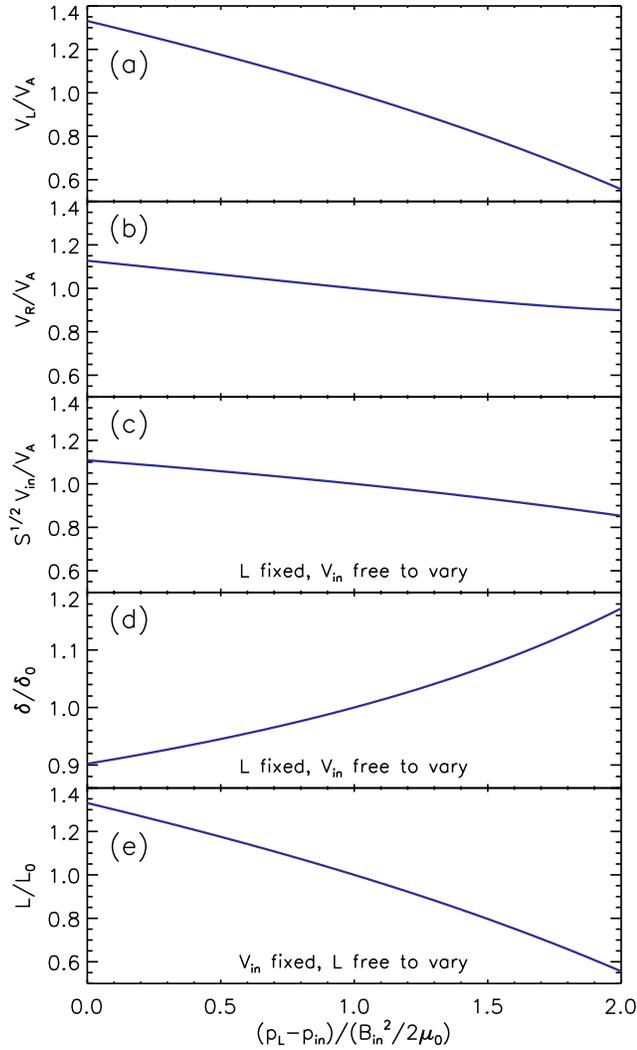}
  \end{center}
  \caption{
    A solution slice along $p_R - p_{in} =B_{in}^2/2\mu_0$ for
    different values of $p_L - p_{in}$ for the incompressible case.
    Shown are 
    (a) $V_L/V_A$, 
    (b) $V_R/V_A$, 
    (c) $S^{1/2}V_{in}/V_A=\sqrt{(V_L+V_R)/2V_A}$, 
    (d) $\delta/\delta_0$, and 
    (e) $L/L_0$.
    Plots (c) and (d) describe the limiting case where $L$ is
    prescribed by the global geometry and that $V_{in}$ is free to
    vary, whereas (e) assumes that $V_{in}$ is prescribed externally
    and that $L$ is free to vary.}
    \label{plot1d}
\end{figure}

Figure \ref{plot1d} shows solutions for the incompressible case as a
function of $p_L-p_{in}$ for fixed $p_R-p_{in} = B_{in}^2/2\mu_0$.
The outflow velocities, calculated using equation (\ref{incomp}) and
shown in Figures \ref{plot1d}a and \ref{plot1d}b, illustrate the weak
dependence that downstream pressure from one side of the current sheet
has on the outflow velocity from the other side of the current sheet.
Figures \ref{plot1d}c and \ref{plot1d}d consider the limiting case
where $L$ is prescribed by external influences on geometry and
$V_{in}$ varies as a function of $p_L$ and $p_R$.  The normalized
reconnection rate, seen in Figure \ref{plot1d}c, changes modestly
despite the large change in $p_L$.  Defining $\delta_0\equiv\sqrt{\eta
L/\mu_0V_A}$, we see that $\delta/\delta_0 = \sqrt{2V_A/(V_L+V_R)}$
increases with $p_L$.  The increased current sheet thickness slows the
reconnection rate slightly by equation (\ref{inflow_whole}).  Figure
\ref{plot1d}e considers a different limiting case for which $V_{in}$
is prescribed due to external driving of reconnection and $L$ varies
as a function of $p_L$ and $p_R$ to maintain the same reconnection
rate.  For this case, $\delta$ is given by equation
(\ref{inflow_whole}) and is independent of downstream pressure.
Defining $L_0\equiv \eta V_A/\mu_0V_{in}^2$, the normalized length is
given by $L/L_0 = (V_L+V_R)/2V_A$.  Figure \ref{plot1d}e shows that
greater downstream pressure reduces the length of the current sheet
for this case, and hence the throughput of mass, in response to
greater downstream pressure.

\subsection{Internal Structure\label{internal}}

Now that the global quantities associated with a current sheet with
asymmetric downstream pressure can be found, we turn our attention to
the internal structure of the current sheet.  The current sheet is
split into three regions of lengths $L_L$, $L_{M}$, and $L_R$, with
boundaries at the flow stagnation point and magnetic field null as
indicated in Figure \ref{asym_diagram}.  The length $L_{M}$ is the
distance between the magnetic field null and the flow stagnation
point, which we will see need not be zero.  The full length of the
current sheet, $2\ltot$, is given by
\begin{equation}
  2\ltot = L_L + L_{M} + L_R.\label{ltot_def}
\end{equation}
We assume in this section that the fields along each boundary are
describable by approximately uniform values for the upstream fields;
however, this may not be justified when current sheet motion relative
to the upstream fields is important or when a long current sheet
develops in a stratified medium such as the wake behind a CME.

As in the model by \citet{cassak:asymmetric}, the position of the flow
stagnation point is set by conservation of mass.  Evaluating equation
(\ref{mass_int}) for the three sections of the current sheet presented
in Figure \ref{asym_diagram} yields the conservation of mass relations
\begin{eqnarray}  
  \rho_{in}V_{in}L_L \sim \rho_LV_L\delta, \label{massL}\\
  \rho_{in}V_{in}{L_{M}} \sim \rho_nV_n\delta, \label{massM}\\
  \rho_{in}V_{in}\left({L_{M}} + L_R\right) \sim 
  \rho_RV_R\delta. \label{massR}
\end{eqnarray} 
where $\rho_n$ is the density and $V_n$ is the outflow component of
velocity at the magnetic field null.  The location of the flow
stagnation point can be derived from equations (\ref{ltot_def}),
(\ref{massL}), and (\ref{massR}), and is given by the relations
\begin{eqnarray}
  L_L \sim 2\ltot \left(\frac{\rho_LV_L}{\rho_LV_L+\rho_RV_R}\right), 
  \label{stagleft}
  \\
  {L_{M}} + L_R \sim 2\ltot
  \left(\frac{\rho_RV_R}{\rho_LV_L+\rho_RV_R}\right).\label{stagright}
\end{eqnarray}

Evaluating the conservation of energy surface integral given in
equation (\ref{energy_int}) in a similar way yields
\begin{eqnarray}
  V_{in}L_L
  \left[ \alpha p_{in} + \frac{B_{in}^2}{\mu_0} \right]
  + V_{in}\delta  \left( \frac{B_{in}B_s}{\mu_0} \right) \sim
  \hspace{1cm} \nonumber\\ \hspace{1.7cm}
  V_L\delta
  \left[
    \alpha p_L + \frac{\rho_L V_L^2}{2}
  \right], \label{left_E}
  \\
  V_{in}{L_{M}}
  \left[ \alpha p_{in} + \frac{B_{in}^2}{\mu_0} \right] \sim
       \hspace{4cm} \nonumber\\ \hspace{1.7cm}
  V_{in}\delta \left( \frac{B_{in}B_s}{\mu_0} \right) 
  + V_n\delta 
  \left[ \alpha p_n + \frac{\rho_n V_n^2}{2} \right],
  \label{mid_E}
  \\
  V_{in}L_R
  \left[ \alpha p_{in} + \frac{B_{in}^2}{\mu_0} \right]
  + V_n\delta 
  \left[ \alpha p_n + \frac{\rho_n V_n^2}{2} \right] \sim
	      \hspace{0.5cm} \nonumber\\ \hspace{1.0cm}
  V_R\delta 
  \left[
    \alpha p_R + \frac{\rho_RV_R^2}{2}
  \right], \label{right_E}
\end{eqnarray}
where $B_s$ is the vertical magnetic field strength at the flow
stagnation point.  When the magnetic field null and flow stagnation
point are separated, there is a Poynting flux across the flow
stagnation point and a kinetic energy flux across the magnetic field
null.

While, in principle, equations (\ref{left_E})--(\ref{right_E}) can be
solved for $L_L$, $L_M$, and $L_R$, we proceed using an alternate
argument to find the separation between the magnetic field null and
flow stagnation point.  In a steady state, the outflow component of
the momentum equation along $z=0$ reduces to
\begin{equation}
  \rho V_x\frac{\partial V_x}{\partial x} = 
  \frac{B_z}{\mu_0}
  \frac{\partial B_x}{\partial z} - \frac{\partial p}{\partial x},
  \label{momentum_outflow}
\end{equation}
where we assume that the magnetic pressure inside the current sheet is
small compared to magnetic tension.  At the flow stagnation point, the
magnetic tension force must cancel the pressure gradient force in a
steady-state system.  Moreover, the magnetic field null is colocated
with the flow stagnation point in a steady state only when there is no
pressure gradient at the magnetic field null.

To quantify this, define $x_n$ as the position of the magnetic field
null and $x_s$ as the position of the flow stagnation point.  By
definition, $x_s$ is given by $V_x(x_s)\equiv 0$.  Evaluating equation
(\ref{momentum_outflow}) at the flow stagnation point using equation
(\ref{ampere_scale}) gives
\begin{equation}
  \frac{B_z(x_s)}{\mu_0}\frac{B_{in}}{\delta} \sim
  \left. \frac{\partial p}{\partial x}\right|_{x=x_s}
\end{equation}
Using that $B_z(x_n)\equiv 0$, a Taylor expansion to first
order yields
\begin{equation}
  B_z(x_s) \simeq \left(x_s-x_n\right) \left. \frac{\partial B_z}{\partial
      x} \right|_{x=x_n}. 
\end{equation}
Hence, we can approximate the distance between the flow stagnation
point and the magnetic field null for a steady state,
\begin{equation}
  {L_{M}} =
  x_n-x_s \sim 
  \left(
    \frac{\mu_0\delta}{B_{in}}
  \right)
  \left(
    \frac{-\left.\partial p/\partial x\right|_{x=x_s}}{
      \left.\partial B_{z}/\partial x\right|_{x=x_n}}
  \right).\label{eps_predict}
\end{equation}

\section{Comparison to Simulations\label{sim}}

In this section, we test the scaling relations and results derived in
Section 3 using resistive MHD simulations for the configuration of
MRX, but with the geometry straightened from cylindrical to linear.
Asymmetry in the outflow direction develops for the push mode of
operation \citep[see][Figure 3]{yamada:mrx} when one downstream wall
is closer to the current sheet than the other downstream wall.  The
driving process in these simulations constrains the position of the
reconnection layer and thus limits time-dependent motion of the
current sheet.  Except as otherwise noted, the numerical method and
simulation setup are described by \citet{murphy:mrx}.

\subsection{Numerical Method and Problem Setup}

The NIMROD code (Non-Ideal Magnetohydrodynamics with Rotation, Open
Discussion) \citep{sovinec:jcp} has been successfully used to model
reconnection in a variety of geometries
\citep[e.g.,][]{hooper:recon,murphy:mrx,zhu:2009}.  NIMROD uses a
finite element expansion in the poloidal plane and, in
three-dimensional simulations, a Fourier representation for the
out-of-plane direction.  The system of equations solved by NIMROD for
the two-dimensional simulations reported in this section are
\begin{eqnarray}
  \frac{\partial\mathbf{B}}{\partial t} = -\nabla \times 
  \left(\eta
    \mathbf{J}-\mathbf{V}\times\mathbf{B}\right)
  +\kappa_{divb}\nabla\left(\nabla\cdot\mathbf{B}\right),
  ,\label{nim_farohms}
\\
  \mu_0\mathbf{J}=\nabla\times\mathbf{B},
\\
  \rho\left(\frac{\partial\mathbf{V}}{\partial t}+\mathbf{V}\cdot
  \nabla\mathbf{V}\right)= \mathbf{J}\times\mathbf{B}-\nabla p
  + \nabla\cdot\rho\nu\nabla\mathbf{V},
\\
  \frac{\partial n}{\partial t}+\nabla\cdot\left(n\mathbf{V}\right)=
  \nabla\cdot D \nabla n , 
  \label{nim_continuity}
\\
  \frac{n}{\gamma-1}\left(\frac{\partial T}{\partial t}
  +\mathbf{V}\cdot\nabla T\right) = 
  -\frac{p}{2}\nabla\cdot\mathbf{V}-\nabla\cdot\mathbf{q}
  + Q,\label{nim_energy}
\end{eqnarray}
where the heat source term $Q=\eta J^2 + \nu\rho\nabla \mathbf{V}^T
\mathbf{:} \nabla\mathbf{V}$ includes Ohmic and viscous heating,
isotropic thermal conduction $\mathbf{q}=-n\chi\nabla T$ is used,
$\nu$ is the kinematic viscosity, $\chi$ is the thermal diffusivity,
and $D$ is an artificial number density diffusivity.  Divergence
cleaning is included in equation (\ref{nim_farohms}) to control
divergence error \citep{sovinec:jcp}.  The ratio of specific heats is
given by $\gamma=5/3$, which corresponds to $\alpha\equiv
\gamma/(\gamma-1)=2.5$.  The resistivity is given by
$\eta/\mu_0=20\mbox{~m}^2\mbox{~s}^{-1}$, which corresponds to a
Lundquist number of $S\sim 100$ using the half-length of the current
sheet and the immediately upstream magnetic field strength and
density.  The magnetic Prandtl number is given by $\mathrm{Pm}\equiv
\nu/(\eta/\mu_0) = 0.25$, with
$\nu=D=\chi=5\mbox{~m}^2\mbox{~s}^{-1}$.  Reconnection is driven by
applying an electric field on the surfaces of two flux cores which
have a minor radius of $9.4$ cm and are located at
$(x,z)=(0\mbox{~cm}, \pm 27\mbox{~cm})$ using the coordinate system
established in section \ref{lineargeom}.  In linear geometry, the flux
cores are infinite cylinders.  The upper and lower boundaries are at
$z=\pm 62$ cm.  For one run, the downstream boundaries are at
$x=-22.5$ cm and $x=32.5$ cm (case A).  For the another run, the
downstream boundaries are at $x=-17.5$ cm and $x=32.5$ cm (case B).

\subsection{Simulation Results\label{linsim}}

\begin{figure}
  \includegraphics[scale=1]{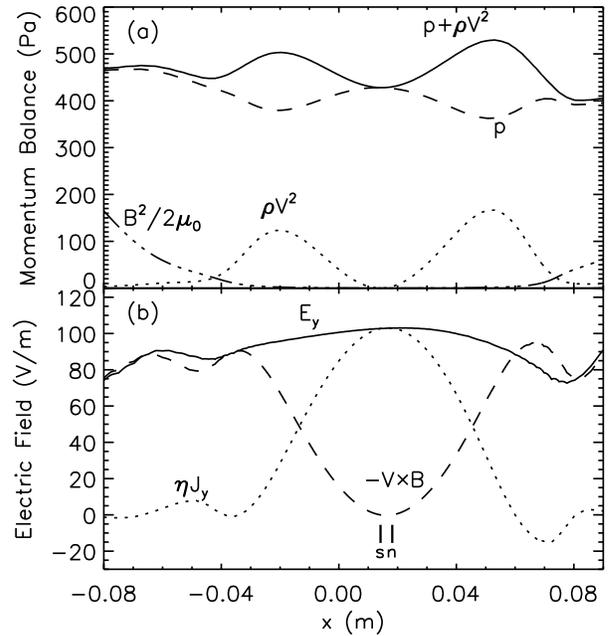}
  \caption{A slice in the outflow direction 
    along $z=0$ of terms from (a) momentum balance and
    (b) electric field balance for case B at 
    $t=13.3\mbox{~}\mu\mbox{s}$.  Vertical bars represent the
    positions of the flow stagnation point (marked `s') and the magnetic
    field null (marked `n').}
  \label{Z0_slice}
\end{figure} 

\begin{figure*}
  \includegraphics[scale=1]{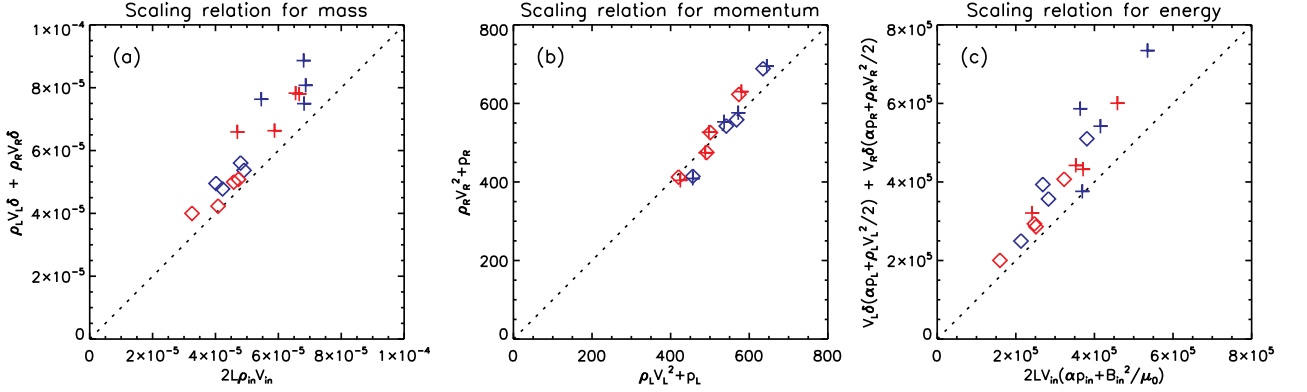}
  \caption{Comparisons between the scaling relationships derived in
    section \ref{lineargeom} and the simulation results.
    Shown in SI units are the left and right hand sides of equations 
    (\ref{mass_whole})--(\ref{energy_whole}) representing scaling
    relations for (a) mass, (b) momentum, and (c) energy.  
    The data points representing cases A and B are plotted in blue and
    red, respectively, for $f=\e^{-1}$ (diamonds) and 
    $f=\e^{-2}$ (plus signs).  The data were extracted at 
    9.1 $\mu$s, 11.2 $\mu$s, 13.3 $\mu$s, and 15.4 $\mu$s.
    The dotted line represents a one-to-one correspondence between the
    left and right hand sides of each scaling relation.  
  }  
  \label{linMRX_LHSRHS}
\end{figure*}

Next, we present results from these simulations and compare them to
the model presented in this paper.  A cut along $z=0$ from case B at
13.3 $\mu$s is shown in Figure \ref{Z0_slice}.  At this time, the flow
stagnation point is at $x=1.42\mbox{~cm}$ and the magnetic field null
is at $x=1.82\mbox{~cm}$, indicating a short separation between the
two points.  Over most of the simulated time the flow stagnation point
is located closer to the side with the impeded outflow than the
magnetic field null in qualitative agreement with equation
(\ref{eps_predict}).  Magnetic pressure is not important within the
current sheet.

\begin{figure*}
  \begin{center}
    \includegraphics[scale=1]{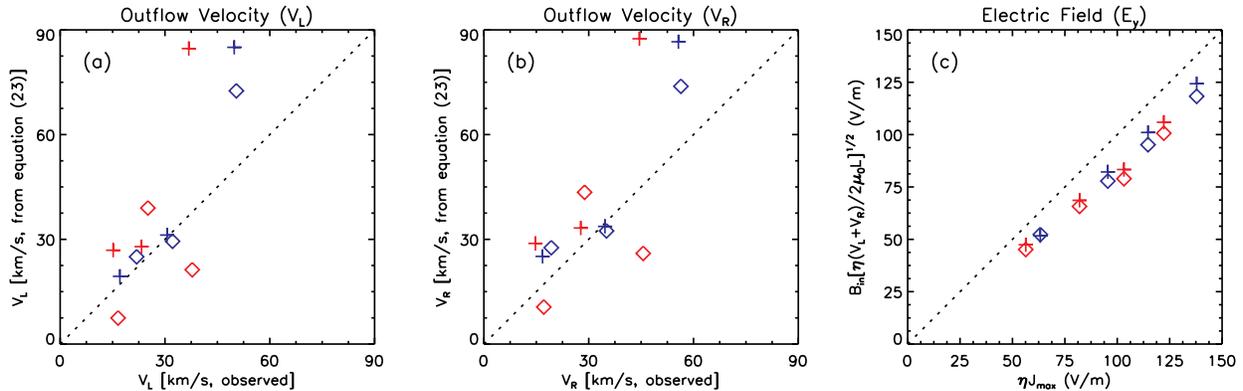}
  \end{center}
  \caption{
    Comparisons between model predictions and simulation results.
    Shown are (a) the outflow velocity $V_L$ compared against equation
    (\ref{linear_cubic}), (b) the outflow velocity 
    $V_R$ compared against equation (\ref{linear_cubic}), and (c) the
    electric field $E_y$ compared against equation (\ref{reconEy}).  
    The symbol usage is described in Figure \ref{linMRX_LHSRHS}.
    \label{VL_VR_E}}
\end{figure*}

To perform quantitative comparisons with theory, the relevant
quantities must be extracted from the numerical results.  The full
length $2\ltot$ of the current sheet is taken to be the distance along
$z=0$ between the two locations where the out-of-plane current density
drops to a fraction $f$ of its peak value, where $f$ is either $1/\e$
or $1/\e^2$.  The thickness of the current sheet, $\delta$, is taken
to be the distance in the $z$ direction between the location where the
out-of-plane current density peaks and where it falls off to $f$ of
its peak value.  The values for the upstream fields are extracted from
the simulation at $z=\pm\delta$ above and below where the current
density peaks.  This method slightly but systematically underestimates
the upstream magnetic field strength.  The values for the downstream
fields are taken at $z=0$ where the out-of-plane current density falls
to $f$ of its peak value.

Comparisons between simulation and our scaling relations are shown in
Figure \ref{linMRX_LHSRHS}, using both of the aforementioned values of
$f$ for both case A and case B. Figure \ref{linMRX_LHSRHS}a compares
the left and right hand sides of equation (\ref{mass_whole}) which
approximates conservation of mass, Figure \ref{linMRX_LHSRHS}b
compares the left and right hand sides of equation
(\ref{momentum_whole}) regarding momentum balance, and Figure
\ref{linMRX_LHSRHS}c compares the left and right hand sides of
equation (\ref{energy_whole}) which approximates conservation of
energy.  Verification of these scaling relations requires that the
data reasonably fit a straight line through the origin.  In Figures
\ref{linMRX_LHSRHS}a, \ref{linMRX_LHSRHS}b, and \ref{linMRX_LHSRHS}c,
we see that the left and right hand sides of each of the equations
approximately fit straight lines through the origin with slopes close
to unity.  In Figures \ref{linMRX_LHSRHS}a and \ref{linMRX_LHSRHS}c,
the slope is slightly greater than unity ($\sim$1.15--1.2).

\begin{figure}
  \begin{center}
    \includegraphics[scale=1]{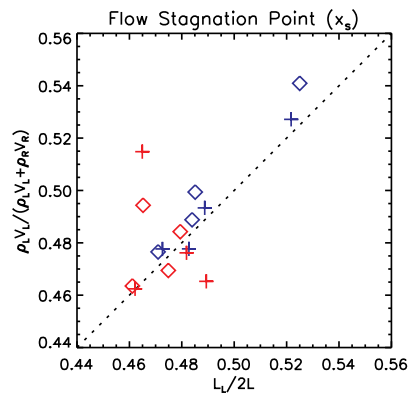}
  \end{center}
  \caption{The position of the flow stagnation point observed in
  simulation compared against the relation given by equation
  (\ref{stagleft}).
  The symbol usage is explained in Figure \ref{linMRX_LHSRHS}.
    \label{xs_compare}}
\end{figure}

Comparisons between the outflow velocities extracted from simulation
and calculated as roots of equation (\ref{linear_cubic}) are shown in
Figures \ref{VL_VR_E}a and \ref{VL_VR_E}b.  Because the positions (and
even existence within the set of real numbers) of roots of high order
polynomials can be sensitive to small changes in the coefficients
\citep[i.e.,][]{wilkinson:1959}, modest differences between the left
and right hand sides of equations (\ref{mass_whole}),
(\ref{momentum_whole}), or (\ref{energy_whole}) sometimes lead to
large errors in the solution for $V_L$ and $V_R$ or the relevant root
becoming complex.  Because of this property common among high order
polynomials, not all of the instances considered have real roots and
there is increased scatter in Figures \ref{VL_VR_E}a and
\ref{VL_VR_E}b beyond what is seen in Figure \ref{linMRX_LHSRHS}.
Despite this, the simulation results show reasonable agreement for
instances where the roots of the polynomials are not greatly impacted
by scaling factors that are not unity in equations (\ref{mass_whole}),
(\ref{momentum_whole}), and (\ref{energy_whole}).  The expression for
the reconnection electric field strength given by equation
(\ref{reconEy}) is compared against simulation in Figure
\ref{VL_VR_E}c, showing good agreement despite a small underprediction
of $\sim$10--20\%.  The positions of the flow stagnation point given
by equations (\ref{stagleft}) and (\ref{stagright}) are tested against
simulation in Figure \ref{xs_compare}.  Despite some outliers, most of
the data points show a good correspondence between the model
predictions and the simulation results.  The scatter is primarily due
to time-dependent effects and the non-uniformity of the upstream
fields.  However, the presence of a local pressure maximum near the
flow stagnation point and magnetic field null complicates the
determination of ${L_{M}}$ for most cases because the pressure
gradient varies significantly in this region; consequently, equation
(\ref{eps_predict}) does not reliably predict the separation between
the flow stagnation point and magnetic field null for these cases.


As a final check for the assumptions of this model, the surface
integrals in equations (\ref{mass_int}), (\ref{momentum_int}), and
(\ref{energy_int}) are calculated along the current sheet boundaries
using the finite element basis functions to interpolate the data.
Evaluating the conservation of mass integral in equation
(\ref{mass_int}) shows that the mass influx is within $\sim$10--25\%
of the mass efflux, indicating modest time-dependence.  Evaluating the
conservation of energy integral given in equation (\ref{energy_int})
shows that the contribution from the term proportional to plasma
pressure is the largest for both the upstream and downstream
boundaries.  During the early stages of reconnection the Poynting flux
out of the layer is comparable to the kinetic energy efflux, but as
reconnection continues to develop the Poynting flux becomes small
($\lesssim$15\%) compared to the kinetic energy efflux.  Evaluating
the outflow component of the conservation of momentum integral given
in equation (\ref{momentum_int}) again shows that the plasma pressure
term is dominant.  Early in time, the downstream magnetic pressure due
to the vertical magnetic field is comparable to the momentum flux out
of the layer but becomes small in comparison as reconnection develops
and the outflow velocities increase with time.  Magnetic tension
forces associated with the upstream boundaries are of the same order
as the momentum flux exiting each side of the layer but are smaller
than the contribution from terms proportional to pressure.  The
tension forces towards each downstream region are symmetric to within
$\sim$5--30\% for case A, but for case B, the tension force directed
towards the obstructing wall is $\sim$2--3 times larger than the
tension force directed towards positive $x$.  The full evaluation of
these surface integrals shows that equations (\ref{mass_whole}) and
(\ref{energy_whole}) representing conservation of mass and energy can
be used to successfully describe the scaling of steady magnetic
reconnection with asymmetry in the outflow direction.  For modest
aspect ratio current sheets such as those associated with the Earth's
magnetotail or spheromak merging, contributions to tension along the
boundary can be important for momentum balance in the outflow
direction and should be considered further in future work.

\section{Summary and Conclusions\label{conclusions}}

Magnetic reconnection with asymmetry in the outflow direction occurs
in many systems in nature and in the laboratory, including planetary
magnetotails, coronal mass ejections, flux cancellation events,
laboratory reconnection experiments, astrophysical disks, and
magnetized turbulence.  In this paper, we perform a control volume
analysis to describe long and thin current sheets with asymmetric
downstream pressure and test these scalings using resistive MHD
simulations of driven reconnection.

In section \ref{lineargeom}, we derive a set of scaling relationships
which describe steady-state magnetic reconnection in a current sheet
with asymmetry in the outflow direction without explicitly specifying
the dissipation mechanism.  We derive expressions for the outflow
velocity for both the compressible and incompressible cases that do
not directly depend on the dissipation mechanism.  When resistive
dissipation is assumed, we present an expression for the reconnection
rate that depends on the outflow velocities from both sides of the
current sheet.  Together, these relations show how the outflow
velocities and reconnection rate depend on a combination of upstream
and downstream parameters.  In the presence of asymmetric downstream
pressure, it is possible to have one Alfv\'enic jet and one
sub-Alfv\'enic jet rather than two bidirectional Alfv\'enic jets.  The
reconnection rate is greatly reduced only when outflow from both sides
of the current sheet is blocked.  This helps explain results by
\citet{oka:asym}, who find that the presence of an obstacle on one
downstream side of the current sheet does not greatly impact the
reconnection rate.

In a steady state, the magnetic field null and flow stagnation point
overlap only in the absence of pressure gradient forces at the
magnetic field null.  When there is a pressure gradient, the magnetic
field null is located on the side of the flow stagnation point which
allows magnetic tension to counter the non-electromagnetic forces at
the flow stagnation point.  The position of the flow stagnation point
can be estimated using conservation of mass when the upstream density
and inflow velocity are approximately uniform.  The position of the
magnetic field null relative to the flow stagnation point can be
estimated using a Taylor expansion around the flow stagnation point.

To test the scaling relations derived in this paper, we perform
two-dimensional resistive MHD simulations of driven reconnection using
the setup of MRX in linear geometry.  Asymmetry in the outflow
direction develops because one downstream wall is closer to the
current sheet than the other downstream wall.  The driving mechanism
of MRX constrains the current sheet position between the flux cores
and limits current sheet motion.  Data extracted from this test show
good correspondence with the scaling relations approximating
conservation of mass, momentum, and energy.  The solution of equation
(\ref{linear_cubic}) for outflow velocities shows reasonable agreement
but increased scatter since the roots of high order polynomials can be
sensitive to small errors in the coefficients.  The reconnection
electric field strength and the flow stagnation point position are
well predicted by equations (\ref{reconEy}) and (\ref{stagleft}).  The
position of the magnetic field null is not well predicted by equation
(\ref{eps_predict}) due to the presence of a local pressure maximum
near these two points.  Exact evaluation of the integrals show that
most of the assumptions of the model are met but that there is a
non-negligible contribution due to tension along the boundary of the
current sheet.

Laboratory plasma experiments such as MRX, SSX, and TS-3/4 provide an
excellent opportunity to study the impact of asymmetry on the
reconnection process.  The pull mode of operation in MRX is
well-suited to investigate reconnection with asymmetry in the inflow
direction due to cylindrical geometry effects \citep{murphy:mrx}.
However, effects related to downstream pressure may need to be
incorporated into the scaling relations of \citet{cassak:asymmetric}.
SSX, TS-3/4, and the push mode of operation in MRX can be used to
study the impact of asymmetry in the outflow direction.  The effects
of asymmetry in the outflow direction (including current sheet motion)
can be further studied by \emph{in situ} measurements in the
magnetotail and by observations of solar reconnection phenomena such
as flux cancellation events, chromospheric jets, solar flares, and
coronal mass ejections.

The model developed in this paper assumes steady-state two-dimensional
antiparallel reconnection in a high aspect ratio current sheet.
Refinements or alternatives to this analysis would benefit from the
inclusion of time-dependent effects such as current sheet motion and
plasmoid formation.  Of particular interest are what determines the
rate of X-line retreat as seen in simulations by \citet{oka:asym} and
how the current sheet structure and dynamics are changed due to
current sheet motion.  Three-dimensional effects have the potential to
enhance the ability of plasma to exit the current sheet
\citep[e.g.,][]{lazarian:turbrecon,sullivan:2008,shimizu:2009}.
Future analyses should consider the uneven contribution of magnetic
tension for modest aspect ratio current sheets.  This would allow the
effect noted by \citet{galsgaard:2000}, \citet{galsgaard:2002}, and
\citet{murphy:mrx}, in which asymmetric outflow develops because the
X-point is displaced towards one end of the current sheet, to be
quantified.

\begin{acknowledgments}
  The authors thank Ellen Zweibel, Jennifer Stone, Michiaki Inomoto,
  John Raymond, Ping Zhu, Yi-Min Huang, Michael Shay, Clare Parnell,
  Dmitri Uzdensky, Mitsuo Oka, Joseph Cassinelli, Richard Townsend,
  K.\ Tabetha Hole, John Everett, Seth Dorfman, and Sam Friedman for
  useful discussions.  N.~A.~M.\ and C.~R.~S.\ acknowledge support
  from NSF Grant PHY-0821899 through the Center for Magnetic
  Self-Organization in Laboratory and Astrophysical Plasmas.
  N.~A.~M.\ acknowledges additional support from NASA grant
  NNX09AB17G-R to the Smithsonian Astrophysical Observatory.
  P.~A.~C.\ acknowledges support from NSF grant PHY-0902479.    
  This research has benefited from use of NASA's Astrophysics Data
  System Bibliographic Services.
\end{acknowledgments}

\bibliographystyle{agufull08}
\bibliography{murphy_asym}  

%
%
%
%
%
%
%
%
%
%


%
%

\end{article}




%
%
%
%
%
%



\end{document}